\documentclass[aps,prl,twocolumn,showpacs]{revtex4}

\usepackage{graphicx}
\usepackage{dcolumn}
\usepackage{bm}

\begin{document}

\preprint{APS/123-QED}

\title{Dilution Effects in Two-dimensional Quantum Orbital System}
\author{Takayoshi Tanaka}
\author{Sumio Ishihara}
\affiliation{ Department of Physics, Tohoku University, Sendai 980-8578, Japan}
\date{\today}

\begin{abstract}

We study dilution effects in a Mott insulating state with  
quantum orbital degree of freedom, termed the 
two-dimensional orbital compass model. 
This is a quantum and two-dimensional version of the orbital model 
where the interactions along different bond directions cause frustration between different orbital configurations. 
A long-range correlation of a kind of orbital   
at each row or column, termed the directional order,  
is studied by means of the quantum Monte-Carlo method. 
It is shown that 
decrease of the ordering temperature due to dilution 
is much stronger than that in spin models.  
Quantum effect enhances the effective dimensionality in the system 
and makes the directional order robust against dilution. 
We discuss an essential mechanism of the dilute orbital systems.  
%
\end{abstract}

\maketitle

Recently, 
there is much 
effort to reveal impurity effects on 
correlated electron systems. 
One of the famous examples is the non-magnetic impurity doping into  
the low-dimensional quantum-magnets, such as the spin-Peierls systems and 
even-leg spin ladders, where the antiferromagnetic (AF) long-range order (LRO) 
is induced by breaking the quantum  spin-singlet correlation. 

Randomly dilute correlated electron systems 
with orbital degree of freedom is another 
class of disordered systems. 
Since this degree of freedom corresponds to 
anisotropy of the electronic clouds, 
the orbital interaction in a solid is essentially directional. 
That is to say, 
Hamiltonian explicitly 
depends on the bond direction unlike spin models \cite{ishihara,kugel}. 
There are competition and frustration 
between the different-bond interactions 
which reduce the effective dimensionaliry in the system. 
Dilution effects in such a unique system have been studied experimentally 
in KCu$_{1-x}$Zn$_{x}$F$_3$ \cite{murakami}. 
One of the doubly degenerate $e_g$ orbitals in a Cu ion  
is occupied by a hole 
and exhibits LRO in KCuF$_3$.  
With substitution of Cu by Zn, 
the orbital ordering temperature $T_{\rm OO}$ rapidly decreases and 
disappears around $x=0.5$. 
This decrease of $T_{\rm OO}$ is much stronger than 
that in dilute magnets and 
in the percolation theory. 

This experimental observation was   
well explained by the calculation based on the orbital Hamiltonian \cite{tanaka}: 
\begin{equation}
{\cal H}= 2J\sum_{i, l}
\left ( {\bf T}_{{\bf r}_i}  \cdot \hat m^l \right )  
\left ( {\bf T}_{{\bf r}_i+\hat l}  \cdot \hat m^l \right ) 
\varepsilon_{{\bf r}_i} \varepsilon_{{\bf r}_i+\hat l}  ,  
\label{eq:eg}
\end{equation}
with a factor $\varepsilon_{{\bf r}_i}$ taking one for Cu and zero for Zn, 
and the bond direction $l$. 
Here ${\bf T}_{{\bf r}_i}$ is the pseudo-spin (PS) operator at site $i$, 
and $\hat m^l$ is a unit vector in the $T^x-T^z$ plane. 
When we take 
$\hat m^l=\left [ \sin\left (  \frac{2n_l}{3}\pi \right ) , \cos \left ( \frac{2n_l}{3} \pi\right ) \right ] $
with $(n_{x},n_{y},n_{z})=(2,1,0)$, 
this model represents the interaction between the $e_g$ orbitals. 
Essence of the orbital dilution 
is frustration between the different-bond interactions 
and a macroscopic orbital degeneracy in the classical ground state 
due to the interactions. 

In this paper, 
we study dilution effects in the two-dimensional and 
quantum version of Eq.~(\ref{eq:eg}) 
with $\hat m^x=\left [ 1, 0 \right]$ and $ \hat m^z=\left [0,1 \right ] $, 
termed the quantum orbital-compass model (OCM) \cite{mishra,nussinov}. 
This is a minimal orbital model representing 
the directional and frustrated interaction between orbitals. 
One of the major different points from the $e_g$-orbital model 
is that the Hamiltonian is invariant 
under a certain local transformations of the PS operator. 
Due to this symmetry, 
degeneracy of the orbital state is 
more serious than that in the $e_g$-orbital model, 
and the conventional site-diagonal order is not expected. 
Instead,  
the long-range correlation for  
one of the PS components appears along the vertical or horizontal direction in a square lattice,  
termed the directional order (DO). 
Dilution effects in DO are investigated 
by means of the quantum Monte-Carlo (QMC) method 
in addition to some rigorous analytical considerations. 
We, in particular, focus on the quantum effects in the orbital dilution. 
It is shown that quantum fluctuation from the classical DO state enhances the effective dimensionality 
in the system and makes DO robust against dilution.  

The explicit form of OCM is given in a two-dimensional square lattice as 
\begin{eqnarray}
{\cal H}= 2J\sum_{i , l=(x,z)} 
T_{{\bf r}_i}^{l} T_{{\bf r}_i+\hat l}^{l} \varepsilon_{{\bf r}_i} \varepsilon_{{\bf r}_i+\hat l}  . 
\label{eq:comps}
\end{eqnarray}
Since the sign of the interaction $J$ is gauged away by rotating ${\bf T}_{{\bf r}_i}$ with respect to the $y$ axis, 
we take it negative. 
This Hamiltonian without dilution ($\varepsilon_{{\bf r}_i}=1$ for $\forall i$) 
is invariant under the following two-symmetry operations \cite{nussinov,dorier};  
(i) The global four-fold symmetry: 
the pseudo-spins at all sites and a crystal lattice are rotated by $\pi/2$, 
simultaneously, 
with respect to the $y$ axis. 
(ii) The local symmetry at each column and row: 
the $z$ ($x$)-component of all PSs 
at each column (row) along the $z$ ($x$) axis are flipped, i.e. 
$T_{r_x, r_z}^{z(x)} \rightarrow -T_{r_x, r_z}^{z(x)}$ for each $r_x\ (r_z)$. 
In addition, in the classical ferromagnetic-like ground state, 
there is the following continuous symmetry: 
(iii) the global PS direction is arbitrary in the $T^x-T^z$ plane. 

Let us explain in more detail the symmetry (ii) which 
is unique in OCM.  
This symmetry operation at a row $r_z$ is done by 
the operator $P_{r_{z}}( \equiv \prod_{r_{x}}\sigma_{r_{x},r_{z}}^{z}$), 
with the Pauli matrices ${\bf \sigma}_{{\bf r}}$, as 
$P_{r_{z}}^{-1}\sigma_{r_{x},r_{z}}^{x}P_{r_{z}}=-\sigma_{r_{x},r_{z}}^{x}$ and 
$P_{r_{z}}^{-1}\sigma_{r_{x},r_{z}}^{z}P_{r_{z}}=\sigma_{r_{x},r_{z}}^{z}$. 
On an equal footing, the symmetry operation at a column $r_x$ is 
done by the operator 
$Q_{r_{x}}=\prod_{r_{z}}\sigma_{r_{x},r_{z}}^{x}$. 
The operators $P_{r_{z}}$ and $Q_{r_{x}}$ commute with ${\cal H}$, 
and $P_{r_{z}}$s commute with each other as well as $Q_{r_{x}}$s, 
but $[Q_{r_{x}}, P_{r_{z}}] \neq 0$. 
The energy eigen-values are characterized by 
the eigen-value set of $P_{r_{z}}$s ($p_{1}, \cdots ,p_{L}$) 
or that of $Q_{r_{x}}$s ($q_{1}, \cdots ,q_{L}$) in a $L \times L(=N) $ square lattice.
That is, at least, there are the $L$ conserved quantities. 
Since $P^2_{r_i}=1$ and $P_{r_i}^\dagger=P_{r_i}$, $p_i=\pm 1$. 
It is easily shown that 
the conventional site-diagonal order parameter 
$N^{-1} \sum_{\bf r} \langle {\bf T}_{\bf r} \rangle$ 
is not invariant under this local-symmetry operation.   
According to the generalized Elitzur's theorem, 
it is shown that the site-diagonal orbital LRO does not occur \cite{batiste}.
Instead, 
a certain kind of LRO, termed the directional order, appears. 
This is a long-range correlation of $T^{z}$ ($T^{x}$) at each column (row), 
and corresponds to a breaking of the global four-fold symmetry (i) explained above. 
This nematic-like order has been studied by means of  
the correlation function 
$C^{zz}_{r_x}\equiv \lim_{r_z \rightarrow \infty} \langle T_{r_x 0}^z T^z_{r_x r_z} \rangle$ 
at zero temperature~\cite{dorier}
and  
$q=N^{-1}\sum_{\textbf{r}} \left ( T_{\textbf{r}}^{z2}-T_{\textbf{r}}^{x2} \right )$ 
in the classical OCM \cite{mishra}.  
We introduce, as an order parameter of DO, a more convenient form $D$ defined by 
\begin{eqnarray}
D&=&N^{-1}(1-x)^{-1}
\nonumber \\
&\times &
\sum_{\textbf{r}}
\big(
T_{\bf{r}}^z T_{\bf{r}+\hat{z}}^z \varepsilon_{\bf{r}} \varepsilon_{\bf{r}+\hat{z}} -
T_{\bf{r}}^x T_{\bf{r}+\hat{x}}^x \varepsilon_{\bf{r}} \varepsilon_{\bf{r}+\hat{x}}
\big) . 
\label{eq:d}
\end{eqnarray}
This parameter $D$ indicates the correlation of $T^{z}$ along the $z$ axis 
or that of $T^x$ along $x$. 
Unlike $q$, this is not a constant for the quantum PS operators. 
Since $D$ commutes with $P_{r_{z}}$ and $Q_{r_{x}}$, being different from ${\bf T}_{\bf r}$, 
it is possible for $\langle D \rangle$ to be finite. 

To examine OCM numerically with and without vacancies, 
we adopt QMC method in a finite-size cluster. 
There is no negative-sign problem in OCM. 
The QMC calculations have been performed in a square lattice of 
$ L\times L$ sites ($L$=14$\sim$18) with the periodic-boundary condition. 
We chose the Trotter number $n$ to be 12$\sim$22, 
and extrapolate the calculated results to $n = \infty$. 
At low temperatures of $T < 0.04J$, we perform the simulation with $n=$60$\sim$72.  
We take, at maximum, 10,000MC steps for measurement 
after 3,000MC steps for thermalization. 
Physical quantities are
averaged over 20$\sim$60MC samples at each parameter set. 
We have checked by MC calculations with $10^6$MC steps, 
which is much longer than the integrated autocorrelation time
determined by the binning analyses, that 
the order parameter values in the present calculation are reliable 
within numerical errors.  
The ordering temperature ($T_{\rm DO}$) of DO is 
estimated from the finite-size scaling analyses 
of the Binder cumulant for $D$ 
defined by $Q=1-\frac{\langle D^{4} \rangle}{3\langle D^{2} \rangle^{2}}$ \cite{binder,mishra}.
At $T=T_{\rm DO}$, a value of $Q$ should be size independent.

\begin{figure}[tb]
\begin{center}
\includegraphics[width=1.00\columnwidth,clip]{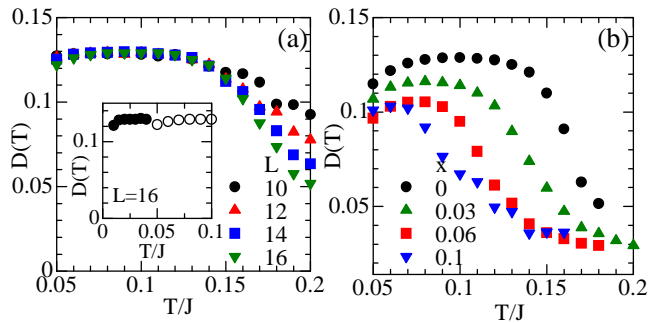}
\caption{(color online). 
(a) Temperature dependence of the directional order parameter ${\overline D}(T)$ 
at $x=0$ for various system sizes $L$. 
The inset shows ${\overline D}(T)$ at low temperatures 
obtained in the calculation with the large Trotter numbers (filled circles). 
Open circles are the same data with those in the main panel. 
(b)  ${\overline D}(T)$ for various impurity concentrations.  
}
\label{fig-2}
\end{center}
\end{figure}
First we show the numerical results in OCM without vacancies.  
Temperature dependence of $\sqrt{\langle D^2 \rangle } (\equiv {\overline D}(T)) $ for several $L$ is 
presented in Fig.~1(a). 
Around $T/J=0.17$, ${\overline D}(T)$ grows with decreasing $T$, 
and is saturated to about 0.13 below $T/J = 0.12$. 
With increasing $L$, 
this dependence becomes steep, 
but the saturated value of ${\overline D}(T)$ at low temperatures does not change much. 
Below $T/J=0.07$, 
${\overline D}(T)$ seems to decrease. 
We have performed careful numerical calculations with the 
larger Trotter numbers $n=$60$\sim$72.  
As shown in the inset of Fig.~1(a), 
${\overline D}(T)$ takes about 0.125 down to $T/J=0.015$. 
Therefore, the reduction of ${\overline D}(T)$ 
in low temperatures is an artifact due to the small Trotter number. 
Temperature dependence of the Binder cumulant $Q$ is presented in Fig.~2. 
The $Q-T$ curves for the several system size $L$ 
cross around $T/J=0.15$ corresponding to the temperature 
where ${\overline D}(T)$ grows. 
We have also calculated the 
susceptibility $\chi_D$ of $D$ and the specific heat $C$. 
Both of them show peak structures around $T/J=0.15$, 
which become steep with increasing $L$. 
We conclude in the present calculation that 
DO is realized in the quantum OCM  
and $T_{\rm DO}=0.150 \pm 0.003$. 
It is noted that the saturated value of ${\overline D}$ at low temperatures is 
about a half of the classical value of $1/4$.  
This reduction is much stronger than that in 
the quantum AF Heisenberg model with $S=1/2$ in a square lattice 
where the saturated staggered moment is about 60$\%$ of $1/2$ \cite{manousakis}. 
We also calculate the site-diagonal order parameter 
$\overline M \equiv \sqrt{\langle {\bf M}^2 \rangle }$ with 
${\bf M}=N^{-1} \sum_{\bf r } {\bf T}_{\bf r}$. 
The calculated $\overline M$ is almost temperature independent and 
its magnitude decreases with increasing $L$. 
This is, 
the site-diagonal order 
does not tend to occur because of the local symmetry (ii), mentioned previously. 
This situation is in contrast to the $e_g$-orbital model where 
the site-diagonal order is realized by 
the order-by-disorder mechanism. 

\begin{figure}[tb]
\begin{center}
\includegraphics[width=0.6\columnwidth,clip]{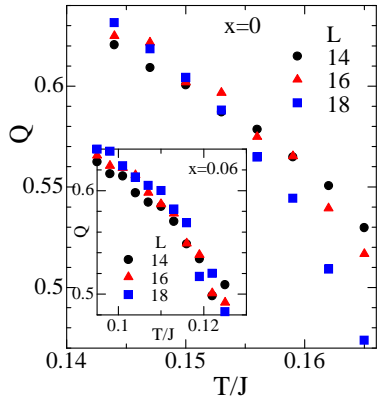}
\caption{(color online). 
Temperature dependence of the Binder cumulant $Q$ at $x=0$ for various system sizes $L$. 
Inset shows temperature dependence of $Q$ at $x=0.06$. 
}
\label{fig-3}
\end{center}
\end{figure}
Now we present the dilution effects in OCM. 
${\overline D}(T)$ 
for several $x$ are shown in Fig.~1(b). 
${\overline D}(T)$ gradually increases with decreasing $T$ and 
is saturated, for example, to about 0.1 at $x=0.06$. 
The temperature where ${\overline D}(T)$ 
rises up is reduced with doping. 
The saturated value of ${\overline D}(T)$ becomes small by doping 
in spite of a factor $(1-x)^{-1}$ in Eq.~(\ref{eq:d}). 
This phenomenon was also observed in the dilute $e_g$-orbital model~\cite{tanaka}
and will be discussed in more detail later. 
The analyses for $Q$ work well at $x=0.06$, as shown in the inset of Fig.~2, 
where $T_{\rm DO}/J$ is identified to be $0.113\pm 0.005$. 
We have checked that, at $x=0.25$,  
the $Q-T$ curves for several $L$ do not cross with each other down to $T=0.01J$. 

The main results in the present work, the $x$ dependence of $T_{\rm DO}$ 
in the quantum OCM, are shown in Fig.~3. 
For comparison, we plot the 
ordering temperature $T_{\rm CL}$ of DO 
in the classical OCM and the Curie temperature $T_{\rm Ising}$ in the $S=1/2$ Ising model in a square lattice. 
$T_{\rm DO}$ decreases monotonically with $x$ and seems to take zero around $x=0.15(\equiv x_c)$.  
It is worth to note that $x_c$ is much smaller than the percolation threshold $0.41$ 
in a square lattice \cite{stauffer}.  
The decrease of $T_{\rm DO}$ is 
much stronger than that in $T_{\rm Ising}$, 
but is weaker than $T_{\rm CL}$. 
In other words, quantum effect makes DO robust against dilution. 

\begin{figure}[tb]
\begin{center}
\includegraphics[width=0.65\columnwidth,clip]{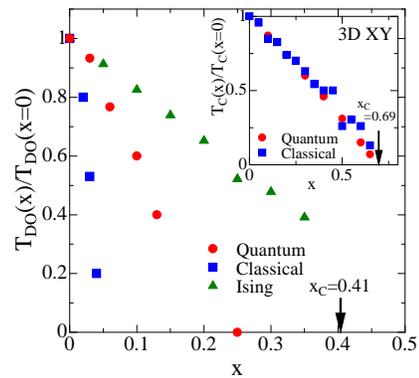}
\caption{
(color online). 
Impurity concentration $x$ dependence of the ordering temperature. 
Closed circles are for $T_{\rm DO}$ in the quantum OCM, 
and closed squares are for $T_{\rm CL}$ in the classical OCM. 
For comparison, the ordering temperature in the Ising model (triangles) are also plotted. 
Bold arrow indicates the percolation threshold $x_c$ in a square lattice.  
The inset shows the ordering temperature $T_{\rm C}$ in the quantum 
XY model (circles) and that in the classical one (squares) in a three-dimensional cubic lattice. 
System size is chosen to be $N=23^3$ at maximum. 
The continuous imaginary-time algorithm is utilized. 
}
\label{fig-1}
\end{center}
\end{figure}
To understand the unusual dilution effects in 
the quantum OCM, we present a rigorous treatment 
of the dilute OCM in two-coupled chains along the $z$ axis, that is, 
a two-leg ladder OCM. 
As mentioned previously, 
the energy eigen-states are classified by 
the eigen-value set $(p_1, \cdots ,p_L)$ of the operator 
$P_{r_z}=\sigma_{r_L r_z}^z \sigma_{r_R r_z}^z $ where 
$r_R (r_L) $ indicates the $x$ coordinates of the right (left) chain in the ladder. 
We consider one of the two-fold ground state with $p_1= \cdots =p_L=1$ 
being degenerate with the state characterized by $p_1= \cdots =p_L=-1$. 
In a row $r_z$ of the ladder, 
the eigen states are $|T^z_{r_L r_z} T_{r_R r_z}^z \rangle=| \uparrow \uparrow \rangle  $ 
and $| \downarrow \downarrow \rangle  $, or their linear combinations 
$|\pm \rangle  \equiv \frac{1}{\sqrt{2}}(|\uparrow \uparrow \rangle  \pm |\downarrow \downarrow \rangle  )$. 
There are relations 
$(T^x_{r_L r_z}T^x_{r_R r_z} ) |\pm \rangle =\pm \frac{1}{4} |\pm \rangle$, 
$T^z_{r_L r_z} |\pm \rangle =\frac{1}{2} |\mp \rangle$ and 
$T^z_{r_R r_z} |\pm \rangle =\frac{1}{2} |\mp \rangle$. 
Therefore, within the subspace of $p_1= \cdots =p_L=1$, 
we introduce new pseuso-spin operators ${\bf S}_{r_z}$  
and show correspondences  
$(T^x_{r_L r_z}T^x_{r_R r_z}) \rightarrow \frac{1}{2}{S_{r_z}^z}$ 
and 
$T^z_{r_L(r_R) r_z} \rightarrow {S_{r_z}^x}$. 
The state $|+ \rangle $ ($|- \rangle $) is the up- (down-) eigen state of the operator $S^z_{r_z}$. 
As a result, 
OCM in this subspace is mapped onto the 
transverse-Ising model in a single chain~\cite{doucot}: 
$
{\cal H}_{eff}=-4J \sum_{r_z }S_{r_z}^x  S^x_{r_z+1}-J\sum_{r_z}S_{r_z}^z. 
$
Now one vacancy is introduced at a site $(r_L, r_z^0)$. 
Since, in the row $r_z^0$, 
we have $P_{r_z^0}=\sigma_{r_R r_z^0}^z $,  
the energy eigen states in this row,   
$| - \uparrow \rangle$ and 
$| - \downarrow \rangle$, belong to the different subspace, 
and $T_{r_R r_z^0}$ is replaced by a $C$-number in each subspace. 
Thus, OCM in the two-leg ladder with one vacancy is denoted by the following model:  
\begin{eqnarray}
{\cal H}_{eff}&=&-4J \sum_{r_z }' S_{r_z}^x  S^x_{r_z+1}-J\sum_{r_z \ne r_z^0 }S_{r_z}^z 
\nonumber \\
&\pm&J(S_{r_z^0+1}^x+S_{r_z^0-1}^x), 
\label{effH}
\end{eqnarray}
where $\sum_{r_z}'$ indicates a sum of $r_z$ except for $r_z^0$ and $r_z^0-1$. 
Signs $+$ and $-$ in the third term correspond 
to the states $|- \uparrow \rangle$ and $| - \downarrow \rangle$ at the site $(r_R, r_z^0)$, respectively.  
This model is the two transverse-Ising chains 
where the external field $\pm J$ along the $S^x$ axis is applied at edges of the chains, i.e. 
the $r_z^0+1$ and $r_z^0-1$ sites . 
In the original language, 
this field acts on $T_{r_R(r_L), r_z^0+1}^z$ and $T_{r_R(r_L), r_z^0-1}^z$, 
and enhances the correlation $\langle T^z_{{\bf r}_i} T^z_{{\bf r}_j} \rangle$  around the vacancy. 
The above analyses accord with an intuitive picture for roles of vacancy in OCM; 
due to the directional interaction in the model,  
orbitals near the vacant site adjust their directions 
to gain the exchange energy. 
In other words, vacancy tends to fix the orbital direction around it.  

\begin{figure}[tb]
\begin{center}
\includegraphics[width=0.65\columnwidth,clip]{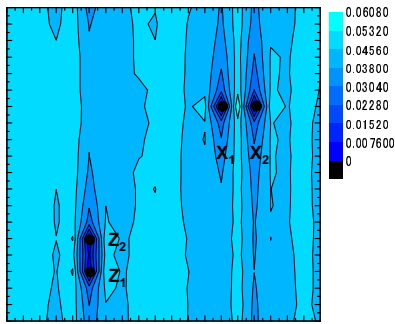}
\caption{
(color online). 
Contour map of local static correlation function $C^{zz}({\bf r}_i)$ 
in a $20 \times 20$ site cluster .
Four vacancies $\rm X_{1(2)}$ and $\rm Z_{1(2)}$ are represented by black circles. 
Temperature is chosen to be $T/T_{\rm DO}(x=0)=$1.27. 
}
\label{fig-4}
\end{center}
\end{figure}
Being based on the above picture, 
we discuss the difference between $T_{\rm DO}$ and $T_{\rm Ising}$ in Fig.~3. 
Consider a situation in a two-dimensional square lattice 
where two vacancies are located closely in a same row, 
e.g. at sites $(r_x-1, r_z)$ and $(r_x+1, r_z)$. 
This probability is of the order of $x$.
Since PS between the vacancies 
has the Ising-type interaction $T_{r_x r_z}^z (T_{r_x r_z+1}^z+T_{r_x r_z-1}^z)$, 
the correlation of $T^z$ is enhanced around the vacancies, as suggested above. 
On an equal footing, 
the correlation of $T^x$ is enhanced around the vacancies 
located closely in a same column. 
It is expected that a certain number of domains/patches, where 
$T^z$ or $T^x$ correlation is enhanced, coexist in a system. 
Competition between such kinds of domains 
reduces the order parameter ${\overline D}(T)$ and the ordering temperature $T_{\rm DO}$. 
To confirm this picture, 
we calculate the local-static PS 
correlation function around the impurity defined by 
$
C^{zz}({\bf r}_i)=\frac{1}{N(1-x)}\sum_{{\bf}_j} \langle T^z_{\bf r_i} T^z_{\bf r_j} \rangle
$ 
(Fig.~4). 
Temperature is chosen to be $T/T_{\rm DO}(x=0)=1.27$. 
Two pairs of vacancies introduced 
in a $20\times20$ site cluster are denoted 
by $X_{1(2)}$ and $Z_{1(2)}$ in the figure. 
It is clearly shown that 
$C^{zz}({\bf r}_i)$ is enhanced between the impurities $\rm X_1$ and $\rm X_2$,
and almost vanishes between $\rm Z_1$ and $\rm Z_2$. 
These results are not seen in dilute spin models. 

Next we focus on quantum effects in dilution, i.e. 
discrepancy between $T_{\rm DO}$ and $T_{\rm CL}$. 
In the classical DO state at $T=0$, 
the system is considered to be decoupled into the independent Ising chains 
along the $x$ or $z$ axis, 
although DO is the two-dimensional order where the four-fold symmetry is broken.   
The rapid decrease of $T_{\rm CL}$ by dilution  
reflects the quasi-one dimensional nature of DO, 
and at finite temperatures, weak two-dimensionality is recovered due to the 
thermal fluctuation.
%
The quantum fluctuation from the classical DO state 
brings about a coupling between the independent chains and increases 
effective dimensionality of DO even at low temperatures. 
This is a kind of resonant states between the independent chains 
and DO becomes robust against dilution. 
This remarkable quantum effect is unique in the orbital system; 
we compare the $x$ dependence of the Curie temperatures $T_{\rm C}$
in the quantum and classical XY models in a cubic lattice (the inset of Fig.~3).  
A magnitude of spin is chosen to be 1/2 in the quantum model. 
It is seen that difference between the two are 
much smaller than that in OCM,  
and both $T_{\rm C}$'s in the XY model seem to disappear at the percolation threshold ($x_c$=0.69). 

Before summarizing the results, we briefly discuss implications of the dilute $e_g$-orbital model. 
In our previous studies~\cite{tanaka}, 
we have suggested that 
a remarkable decrease of $T_{\rm OO}$ is related to the tilting of PS around impurities. 
Roles of this PS tilting become clearer 
by the present analyses of OCM. 
The orbital interaction in the $e_g$-orbital model 
is essentially of the Ising-type along one direction in a cubic 
lattice, and there is frustration between them, as well as OCM. 
Introduced vacancies break the local symmetry and release the frustration. 
According to vacancy configurations, a specific type of the orbital correlation 
is stabilized around vacancies. 
For example, when two vacancies are located closely at a same row along the $z$ axis, 
the orbital correlation of $d_{3x^2-r^2}/d_{y^2-z^2}$ or $d_{3y^2-r^2}/d_{z^2-x^2}$ 
are stabilized around them. 
Competition between these domains/patches with different PS correlations 
reduce the order parameter and 
may  cause the glass-like phase. 

In summary, 
we study the dilution effects in 
a quantum and two-dimensional version of the orbital model, i.e. the orbital compass model, 
where the vertical or horizontal correlation of a PS component, 
termed the directional order, is expected from the local symmetry.   
Reduction of the ordering temperature of DO  
is much stronger than that in the spin model. 
Quantum fluctuation increases the effective dimensionality of the system 
and makes DO robust against dilution. 
Competition between domains with different orbital configurations 
is essence of the dilute orbital system.  

The authors thank to M. Matsumoto for several helpful discussion 
and providing calculation data for the inset in Fig. 3, 
and H. Matsueda for valuable comments. 
This work was supported by Grant-in-Aid for Scientific Research on priority Areas, 
JSPS KAKENHI (16340094, 16104005) and TOKUTEI "High Field Spin Science in 100T" (No.451, 18044001) from MEXT, 
NAREGI, and CREST. 
T. T. appreciates financial support from JSPS.


\begin{thebibliography}{10}

\bibitem{ishihara}
S.~Maekawa,  
{\it et al.}, 
{\it Physics of Transition Metal Oxides}
(Springer Verlag, Berlin, 2004), and references therein.
\bibitem{kugel} 
K. I. Kugel and D.~I. Khomskii, 
Sov. Phys. JETP \textbf{37}, 725 (1973). 
\bibitem{murakami}
N.~Tatami, S.~Niioka, and Y. Murakami, (to be published).
\bibitem{stauffer}
D. Stauffer, Phys. Rep. {\bf 54}, 1 (1979). 
\bibitem{tanaka}
T.~Tanaka, 
{\it et al.}, 
Phys. Rev. Lett. \textbf{95}, 267204 (2005).
\bibitem{mishra}
A.~Mishra, 
{\it et al.}, 
Phys. Rev. Lett. \textbf{93}, 207201 (2004).
\bibitem{nussinov}
Z.~Nussinov, 
{\it et al.}, 
Euro. Phys. Lett. {\bf 67}, 990 (2004). 
\bibitem{dorier}
J. Dorier, 
{\it et al.}, 
Phys. Rev. B \textbf{72}, 024448 (2005). 
\bibitem{batiste}
C.D. Batista and Z. Nussinov, 
Phys. Rev. B. \textbf{72}, 045137 (2005).
\bibitem{binder}
K. Binder, Phys. Rev. Lett. {\bf 47}, 693 (1981).  
\bibitem{manousakis}
E.~Manousakis, 
Rev.~Mod.~Phys. {\bf 63}, 1 (1991). 
\bibitem{stinchcombe}
R.~B.~Stinchcombe, J. Phys. C {\bf 13}, 5565 (1980). 
\bibitem{doucot}
B.~Doucot, 
{\it et al.}, 
Phys. Rev. B {\bf 71}, 024505 (2005). 
%

\end{thebibliography}
\end{document}